\documentclass[aps,nofootinbib,twocolumn,superscriptaddress]{revtex4-1}
\pdfoutput=1
\usepackage[utf8]{inputenc}
\usepackage[pdftex]{graphicx}
\usepackage{amssymb}
\usepackage{amsmath}
\usepackage{color}
\usepackage{bm}
\usepackage[caption=false]{subfig}
\usepackage{placeins}
\usepackage{subfig}
\usepackage{paralist}
\usepackage{tabularx}
\usepackage[english]{babel}

\newcommand{\Cs}{${\rm Cs_2CoCl_4}$}

\RequirePackage[
   hyperindex,colorlinks,bookmarksnumbered,
   plainpages=true, pdfstartview=FitH,bookmarks=false
]{hyperref}
\hypersetup{linkcolor=blue,urlcolor=blue,citecolor=blue}

\definecolor{darkgreen}{rgb}{0,0.5,0}
\definecolor{darkblue}{rgb}{0,0,0.5}
\definecolor{purple}{rgb}{0.35,0,0.35}
\definecolor{orange}{rgb}{1,0.5,0}
\definecolor{wbcolor}{rgb}{0,.6,1}
\definecolor{todocolor}{rgb}{1,0,0}
\definecolor{jvdcolor}{rgb}{0,0,1}

\newcommand{\Eq}[1]{Eq.~(\ref{#1})}

\newcommand{\Sec}[1]{Sec.~\ref{#1}}

\newcommand{\Ref}[1]{Ref.~[\onlinecite{#1}]}

\newcommand{\Fig}[1]{Fig.~\ref{#1}}

\newcommand{\Figs}[1]{Figs.~\ref{#1}}

\newcommand{\RMP}{\textit{Rev. Mod. Phys. }}

\graphicspath{ {Figures/}}

\begin{document}

\title{Dynamic structure factor of 
the spin-$\frac{1}{2}$ XXZ chain in a transverse field}
\author{Benedikt Bruognolo}
\author{Andreas Weichselbaum}
\author{Jan von Delft}
\affiliation{Physics Department, Arnold Sommerfeld Center for Theoretical Physics, and Center for NanoScience,
Ludwig-Maximilians-Universit\"at, Theresienstra{\ss}e 37, 80333 M\"unchen, Germany}
\author{Markus Garst}
\affiliation{Institut f\"ur Theoretische Physik, Universit\"at zu K\"oln, Z\"ulpicher Str. 77a, 50937 K\"oln, Germany}
\affiliation{Institut f\"ur Theoretische Physik, Technische Universit\"at Dresden, 01062 Dresden, Germany}


\begin{abstract}
The spin-$\frac{1}{2}$ XXZ chain with easy-plane anisotropy in a transverse field describes well the thermodynamic properties of the material \Cs\, in a wide range of temperatures and fields including the region close to the spin-flop Ising quantum phase transition.
For a comparison with prospective inelastic neutron scattering experiments on this compound, we present results of an extensive numerical study of its dynamic structure factor $\mathcal{S}^{\alpha \beta}(k,\omega)$ using matrix-product-state (MPS) techniques. Close to criticality, the dynamic part of the correlator $\mathcal{S}^{xx}$ longitudinal to the applied field is incoherent and possesses a small total weight as the ground state is already close to saturation. The transverse correlator $\mathcal{S}^{zz}$, on the other hand, is dominated by a coherent single-particle excitation  with additional spectral weight at higher energies that we tentatively attribute to a repulsively bound pair of particles. With increasing temperature, the latter quickly fades and spectral weight instead accumulates close to zero wave vector just above the single-particle energy. On a technical level, we compare the numerical efficiency of real-time evolution to an MPS-based Chebyshev expansion in the present context, 
finding that both methods yield results of similar quality at comparable numerical costs.
 \end{abstract}

\maketitle


\section{Introduction} \label{sec:intro}

A transverse magnetic field applied to a spin-$\frac{1}{2}$ XXZ chain reduces the remaining U(1) spin-rotation symmetry and immediately results in a gapped ground state whose classical analog corresponds to a spin-flop phase with long-range N\'eel order. Increasing the magnetic field beyond a critical value $H_c$, this long-range order is lost at a Ising quantum phase transition. Such spin chains govern the magnetic properties of the material \Cs\ in a wide regime of temperatures and fields \cite{Algra1976,McElearney1977,Smit1979,Duxbury1981,Kenzelmann2002,Chatterjee2003,Siahatgar2008,Breunig2013,Breunig2015}.
They effectively emerge from spin-$\frac{3}{2}$ Heisenberg chains attributed to Co$^{2+}$ ions whose tetrahedral  environment results in a strong single-ion anisotropy. The latter splits the four levels of each spin-$\frac{3}{2}$ into two doublets, and the low-energy doublet provides an effective spin-$\frac{1}{2}$ degree of freedom. Projecting the Hamiltonian onto this low-energy subspace \cite{Breunig2013,Toskovic2016}, XXZ chains arise with easy-plane anisotropy. The CoCl$_4$ tetrahedra of neighboring chains are tilted with respect to each other which leads to two different easy planes within a single unit cell, so that only a nonstaggered transversal magnetic field can be applied along the crystallographic $b$ axis.
In a recent study \cite{Breunig2013}, it was shown that the thermal expansion and specific heat of \Cs\ below a temperature of approximately 2.5 K and for transverse fields smaller than approximately 3 T can be consistently explained in terms of the spin-$\frac{1}{2}$ XXZ chain Hamiltonian. This parameter range also encompasses the regime of Ising quantum criticality at $\mu_0 H_c \approx 2$ T. At much lower temperatures of approximately 300 mK, the interchain coupling stabilizes three-dimensional long-range order with various different phases as a function of magnetic field \cite{Breunig2015}. 

Whereas neutron diffraction experiments on \Cs\, were carried out already more than ten years ago \cite{Kenzelmann2002}, inelastic neutron scattering studies, as far as we know, have not been performed yet. Such an experiment would access the components of the  dynamical spin-spin correlation functions of the XXZ Hamiltonian in a transverse field,
\begin{equation} \label{eq:DSSF_def}
\mathcal{S}^{\alpha \beta}(k,\omega) = \sum_{j} e^{-i k j}  \Big[ \int\limits_{-\infty}^\infty {\rm d}t \ e^{i \omega t} \langle \hat{S}^{\alpha}_j (t) \hat{S}^{\beta}_0 \rangle \Big],
\end{equation}
where $\hat{S}^{\alpha}_j (t)$ is a spin-$\frac{1}{2}$ operator in the Heisenberg picture with $\alpha = x,y,z$, and the sum extends over sites $j$ of the one-dimensional lattice with unit lattice spacing. The expectation value is taken with respect to the XXZ Hamiltonian
\begin{equation} \label{eq:H12}
\hat H = \sum_j J \Big[ ( \hat{S}_j^x \hat{S}_{j+1}^x + \hat{S}_j^y \hat{S}_{j+1}^y) + \Delta \hat{S}_j^z \hat{S}_{j+1}^z - h \hat{S}_j^x \Big].
\end{equation}
For \Cs\ the parameters were estimated in \Ref{Breunig2013} to be $J/k_B \approx 3$ K 
and $\Delta \approx 0.12$. In the following, we exclusively use this value for $\Delta$ and measure energies in units of $J$. The Ising quantum phase transition then occurs at the dimensionless critical field $h_c \approx 1.56$.

The correlation functions \eqref{eq:DSSF_def} have been theoretically investigated before by Caux, Essler and L\"ow (CEL) \cite{Caux2003} using exact results in combination with a mean-field approximation (MFA). Here, 
we study these correlators numerically with a quasiexact matrix-product-state (MPS) approach as a function of transverse field at zero and finite temperatures $T$, and we extensively compare to the results of CEL. 
In particular, we employ the time-dependent adaption of the density matrix renormalization group (tDMRG) \cite{Feiguin_PRL_2004,Vidal_PRL_2004,Daley_2004} in the MPS framework to carry out the real-time evolution of the real-space correlators in \Eq{eq:DSSF_def} before Fourier transforming into momentum and frequency space. 
The results at finite $T$ are obtained by matrix-product purification \cite{Verstraete2004,Feiguin2005}. For a recent work on the dynamic structure factor of the XXZ chain but with easy-axis anisotropy see \cite{Zhe2015}.

The main findings of our numerical study are the following. The dynamic part of the correlator $\mathcal{S}^{xx}$ longitudinal to the applied field is confirmed to be incoherent close to quantum criticality. Moreover, it possesses a small total weight as the ground state is already close to saturation. The correlator $\mathcal{S}^{zz}$ transverse to the field and longitudinal to the hard axis is dominated by a coherent single-particle excitation close to the critical field in agreement with the findings of CEL. This coherence gets lost with decreasing field as the hybridization with two-particle excitations becomes more and more important. Furthermore, we find additional spectral weight at higher energies that we tentatively ascribe to a repulsively bound pair of particles, which is not anticipated in the MFA of CEL. A finite temperature is expected to destabilize such pairs. Correspondingly, we find that this weight quickly decreases with increasing $T$, and it is redistributed close to zero wave vector just above the single-particle energy.
The interesting and rich physics of repulsively bound particle pairs in the XXZ spin-$\frac{1}{2}$ chain might thus be observable in the spin-spin correlations of the material 
\Cs.

On a technical level, we compare the numerical efficiency of the real-time evolution in the present context to a recently developed MPS-based Chebyshev expansion (CheMPS) \cite{holzner}. Our main conclusion is that CheMPS produces zero-temperature spectral functions of similar quality as tDMRG at comparable computational costs. Accordingly, the CheMPS setup must appropriately deal with a growing amount of entanglement in the MPS to produce reliable results.

The paper is structured as follows.
In \Sec{sec:method} we briefly review the approximation of CEL used in their computation of the correlators \eqref{eq:DSSF_def} and introduce the two matrix-product-state techniques, tDMRG and  CheMPS, employed in our numerical calculations. Our results for the dynamic structure factor are presented in \Sec{sec:results} and compared to the approximation of CEL. 
The paper ends with a short discussion in \Sec{sec:conclusion}. Technical details on tDMRG and CheMPS, including our comparison of their numerical efficiency, are presented in the Appendix.


\section{Methods} \label{sec:method}
\subsection{Approximation of CEL} \label{sec:MFA}
The approximation employed by Caux, Essler, and L\"ow (CEL) \cite{Caux2003} involves two steps. First, 
after a Jordan-Wigner transformation of the Hamiltonian \eqref{eq:H12} the interaction between Jordan-Wigner fermions is treated within a self-consistent mean-field approximation (MFA). This amounts to solving three coupled nonlinear equations numerically.  The validity regime of the MFA was determined by CEL with the help of DMRG calculations of thermodynamic quantities.
In a second step, the structure factor \eqref{eq:DSSF_def} is evaluated with respect to the mean-field Hamiltonian, that can be identified with an effective anisotropic XY spin chain. The spin-spin correlator longitudinal to the magnetic field $\mathcal{S}^{xx}(k,\omega)$ reduces to a density-density correlation function of Jordan-Wigner fermions that can be straightforwardly computed. The spin-spin correlators transverse to the field $\mathcal{S}^{\alpha\beta}(k,\omega)$ with $\alpha,\beta = y,z$, on the other hand, contain Jordan-Wigner strings so that a further approximation is employed. 
Exact results for the XY spin chain are now exploited to approximate the transverse spin-spin correlator either by the 
contribution of the two-particle sector at intermediate fields, $h<h_c$, or by the contribution of the single-particle sector at larger fields, $h>h_c$.

CEL also discuss the range of validity of the MFA by comparing thermodynamic quantities to static  density matrix renormalization group (DMRG) calculations. They conclude (for $\Delta = 1/4$) that the MFA should work well for large fields $h \gtrsim 1.5$ whereas for intermediate field strengths $0.5 \gtrsim h \gtrsim 1.5$ it should provide at least qualitatively correct results. It breaks down however in the low-field limit $h \to 0$.

\subsection{Numerical matrix-product-state techniques}
To capture all facets of the interacting model (\ref{eq:H12}) beyond the approximation of CEL, we employ quasiexact numerical simulations in a matrix-product-states (MPS) setup. The MPS framework offers different approaches to evaluate the components of the dynamic structure factor \eqref{eq:DSSF_def} in frequency space. Here, we mostly use the time-dependent adaption of the density matrix renormalization group (tDMRG) \cite{Feiguin_PRL_2004,Vidal_PRL_2004,Daley_2004} to evolve the real-space spin-spin correlation function in time. The dynamic spin structure factor in frequency space is then obtained by a subsequent Fourier transform of the real-time data. At zero temperature, we start from the ground state of the system obtained with standard DMRG  \cite{White1992,schollwock_2005,schollwock_2011} before applying the local perturbation $\hat{S}^{\beta}_0$ and evolving the state in real time. To obtain finite-temperature correlators, the initial MPS is chosen to be a thermal state representing the purified density matrix at a certain temperature \cite{Verstraete2004,Feiguin2005}.  Details on our tDMRG implementation, the post-processing by means of Fourier transform and the chosen numerical parameters can be found in Appendix \ref{app:tDMRG}. We emphasize that all results presented in \Sec{sec:results} were obtained using tDMRG.

To conclude this section, we briefly mention a point of technical interest for readers with a numerical MPS background. Recently, an MPS-based Chebyshev expansion technique (CheMPS) has been successfully established as a competitive alternative to tDMRG \cite{holzner,braun,Ganahl_PRB_2014,Tiegel_PRB_2014,Wolf_PRB_2014}. It evaluates dynamic correlators directly in frequency space avoiding the Fourier transform required in any real-time approach. However, it still remains unclear which of the two methods, CheMPS or tDMRG, is more efficient for computing spectral functions. To gain some insight into this open question, we conducted a detailed comparison for the present problem at zero temperature. We found that both methods yield results of similar quality at almost identical computational costs. For an extended discussion of technical details of CheMPS, and a comparison of the performance of tDMRG and CheMPS for the present model system, the reader is referred to the Appendices \ref{app:CheMPS} and \ref{app:CheMPSvstDMRG}, respectively.

For completeness, we note that the correction-vector (CV) method can  also be employed to calculate the dynamic structure factor at zero temperature \cite{Ramasesha_1997,Kuehner_PRB_1999,Jeckelmann_PRB_2002,Jeckelmann_2008}. 
However, CV requires individual calculations for each frequency point $\omega$ and is therefore not practicable in the context of this work. In comparison, tDMRG and CheMPS are significantly more efficient since these methods can access the entire frequency axis using a single calculation. 


\section{Results} \label{sec:results}
\subsection{Phase diagram} \label{sec:phasediag}
\FloatBarrier

\begin{figure}[h!]
\centering
\includegraphics[width=\linewidth]{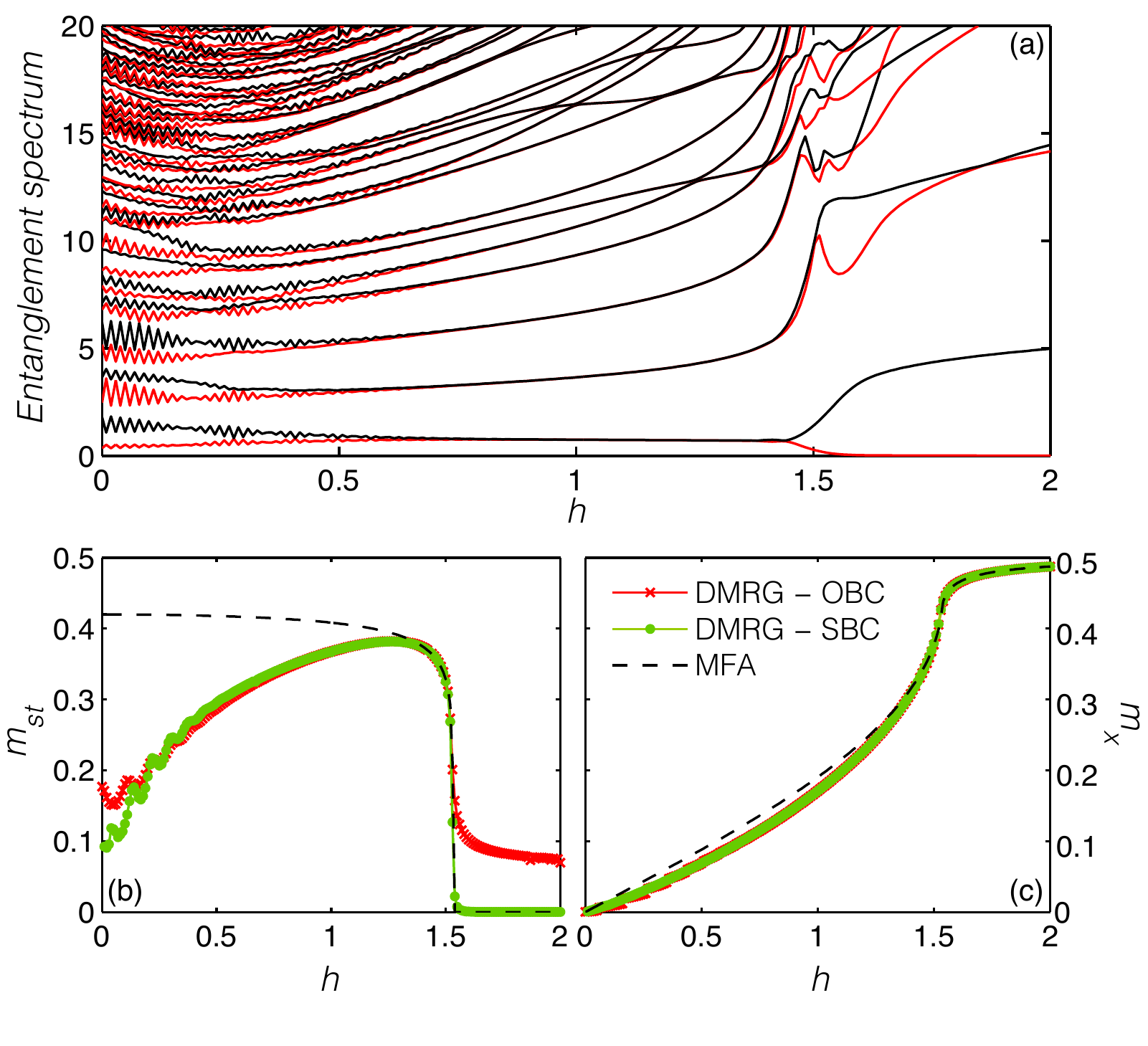}
\vspace{-25pt}
\caption{(a) Entanglement spectrum, (b) staggered per-site magnetization $m_{\rm st}$, and (c) per-site magnetization $m_{x}$ along the in-plane field direction as functions of $h$. To generate the entanglement spectrum, we used a system of $N=301$ spins with a site-dependent field that linearly increases along the chain. Every other state in the entanglement spectrum is shown in red for better visual contrast. The data in panels (b) and (c) was generated from individual DMRG runs for each $h$ on a system with $N=100$ spins using both open and smooth boundary conditions (OBC/SBC). The phase transition from the spin-flop to the spin-polarized phase occurs around $h_c \approx 1.56$ beyond which the order parameter $m_{\rm st}$ vanishes. 
%
}
\label{fig:stagm}
\end{figure}

In order to identify the position of the Ising quantum phase transition of the Hamiltonian \eqref{eq:H12} we have first considered its ground-state properties. The panels in \Fig{fig:stagm} illustrate  distinct static features of the different ground state phases. The data in  panel (a) represents the entanglement spectrum, which is generated from a single ground-state DMRG calculation \cite{White1992,schollwock_2005} of a system with $N=301$ sites while keeping all states associated with singular values larger than $\epsilon_{\rm SVD} = 10^{-5}$. We chose a site-dependent magnetic field $h_j$, which is increased in small steps of $0.01$ throughout the chain from $h_1=-0.5$ at the first site to $h_{301} =2.5$ at the last site. 

This setup provides a quick snapshot of the physics of the different phases vs.~magnetic field along the chain within a single DMRG run and does not require a separate calculation for each value of the magnetic field \cite{Zhu_2013,Zhu_2013b}. While finite-size effects in the bulk part of the chain are reduced in this setup leading to a smooth tuning of the spectrum as a function of $h$, blurred effective finite-size effects are present and depend on the speed of the tuning. In the present case, however, the position of the phase boundary is already in good agreement with the calculations from homogeneous systems in \Figs{fig:stagm}(b) and (c).

By cutting 
the chain on each bond and diagonalizing the reduced density matrix $\hat{\rho}_j$, we obtain the entanglement spectrum $\xi^j_k$ as a function of $h$ from the spectral decomposition $\rho^j_k$ of $\hat{\rho}_j$, i.e., $\xi^j_k = - \log{\rho^j_k}$. The entanglement spectrum displays a smooth behavior in both the spin-flop and the spin-polarized phase and nicely captures the distinct ground-state degeneracy in the two phases. Whereas the ground state is twofold degenerate in the spin-flop phase $0 < h < h_c$, it is unique within the spin-polarized phase, $h>h_c$. 

To locate the critical point quantitatively, we study the order parameter of the system, represented by the staggered magnetization. Since a finite length $N$ breaks translational symmetry, leading to
\begin{equation}
\sum_j \langle \psi_0 | (-1)^j \hat{S}^y_j | \psi_0 \rangle = \sum_j \langle \psi_1 | (-1)^j \hat{S}^y_j | \psi_1 \rangle = 0,
\end{equation}
we calculate the order parameter using

\begin{figure*}[t]
\centering
\includegraphics[width=\linewidth]{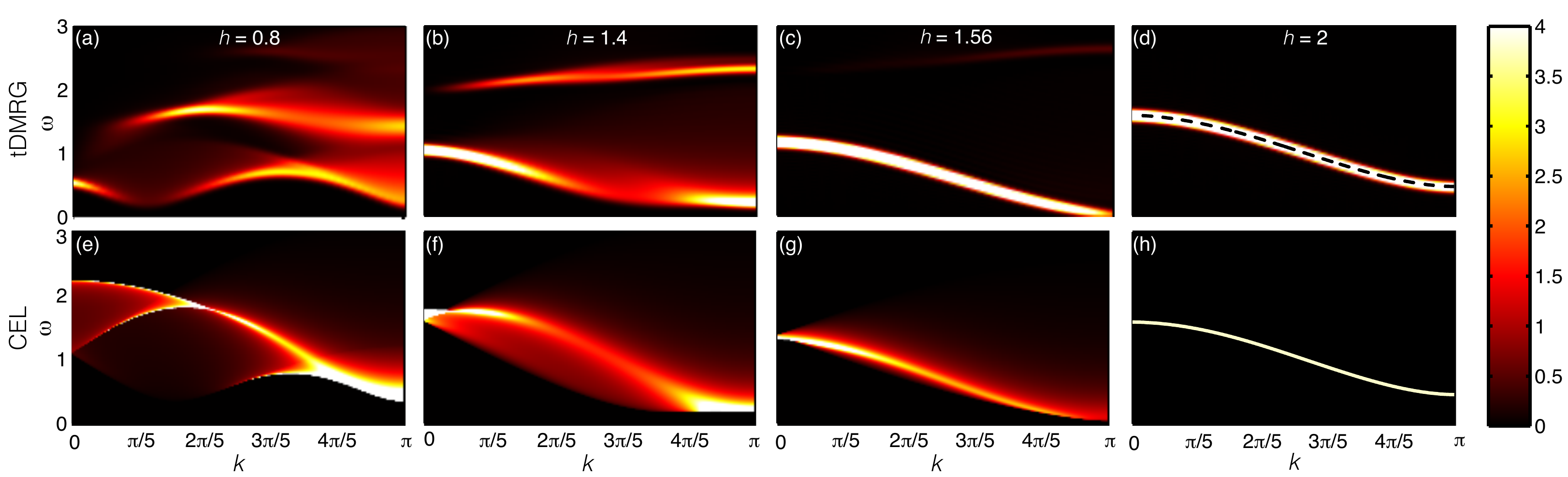}
\vspace{-15pt}
\caption{Dynamic spin structure factor $\mathcal{S}^{zz}(k,\omega)$  longitudinal to the hard axis and transverse to the applied magnetic field $h$, at zero temperature. (a)-(d) show the numerical results of tDMRG [$N=100$, $t_{\rm max}=60$; see Appendix \ref{app:tDMRG_T0} for details] whereas (e)-(h) display the corresponding CEL approximation with contributions from (e)-(g) the two-particle sector and (h) the one-particle sector only. The system is gapped within the spin-flop phase [(a),(b),(e),(f)] with an incoherent spectrum. The gap closes at the quantum phase transition, (c),(g), at $h_c \approx 1.56$. The gap reopens within the spin-polarized phase, (d),(h), where the spectrum is dominated by a single coherent mode [dashed line in (d)] in excellent agreement with the CEL approximation. In panel (b) a distinct higher-energy branch is visible that is absent in the CEL spectra of panel (f).
%
}
\label{fig:DSSF_ZZ}
\end{figure*}

\begin{equation}\label{eq:mstagg}
m_{\rm st} = \frac{1}{N} \sum_j \langle \psi_0 | (-1)^j \hat{S}^y_j | \psi_1 \rangle,
\end{equation}
where $|\psi_0\rangle$ is the ground state and $|\psi_1\rangle$ the first excited state of the system. \Fig{fig:stagm}(b) illustrates the dependence of the order parameter on the in-plane field $h$ using both MFA and ground-state DMRG calculations. Both methods nicely agree for larger fields and pinpoint the critical point at $h_c \approx 1.56\pm0.01$, without performing any further finite-size scaling. 
Since the MFA works poorly for small fields, we observe strong deviations between MFA and DMRG within the spin-flop phase -- a phenomenon which we will reencounter when calculating the components of the dynamic structure factor in \Sec{sec:DSSF}.

We note that the DMRG calculations of $m_{\rm st} $ are plagued by strong finite-size effects when using a standard setup with open boundary conditions (OBC) in the spin-polarized phase, as illustrated by the large finite value of the red curve for $h > h_c$ in \Fig{fig:stagm}(b). The finite-size effects can be significantly reduced for high fields by employing the concept of smooth boundary conditions (SBC) \cite{Vekic1993,Vekic1996} in a small region of 10 sites on the edges of the system (blue curve). The idea of SBC is to smoothly decrease the parameters of the Hamiltonian to zero at both ends of the chain to avoid having a sharp and rigid boundary as in the OBC setup. However, finite-size effects for small fields, albeit reduced with SBC, are not completely absent as indicated by the nonzero value of $m_{st}$ at zero field.

Other quantities such as the magnetization per site, $m_x= \frac{1}{N}\sum_j \langle \psi_0 | \hat{S}_j^x | \psi_0 \rangle$, are already well converged in the OBC setup. As illustrated in \Fig{fig:stagm}(c), a nonzero field immediately leads to a finite magnetization which increases monotonically with $h$. 
Note that even in the spin-polarized phase at $h > h_c$, the magnetization is not saturated yet due to quantum fluctuations. Full saturation is only reached in the limit of infinitely strong magnetic fields.

\subsection{Dynamic structure factors at $T=0$} \label{sec:DSSF}

In the following, we present the numerical tDMRG results for various components of the zero-temperature dynamic structure factor and compare them to the approximation of CEL. Numerical details on our tDMRG implementation can be found in Appendix \ref{app:tDMRG_T0}.
We will discuss the contribution $\mathcal{S}^{zz}$ longitudinal to the hard axis and transverse to the magnetic field, the contribution $\mathcal{S}^{xx}$ longitudinal to the magnetic field, and the spin-flip contribution $\mathcal{S}^{+-} = \mathcal{S}^{xx} + \mathcal{S}^{yy} + i(\mathcal{S}^{yx} - \mathcal{S}^{xy})$. 
For our analysis, we choose four representative values of the magnetic field $h = 0.8, 1.4, 1.56, 2$: the first two are located within the spin-flop phase, the third corresponds to the critical field $h_c$, and the last is located within the polarized phase. We do not consider the limit of zero magnetic field, $h=0$, as the dynamic structure factor in this case is well known \cite{Caux2005,Pereira2006}.

 \begin{figure*}[t]
\centering
\includegraphics[width=\linewidth]{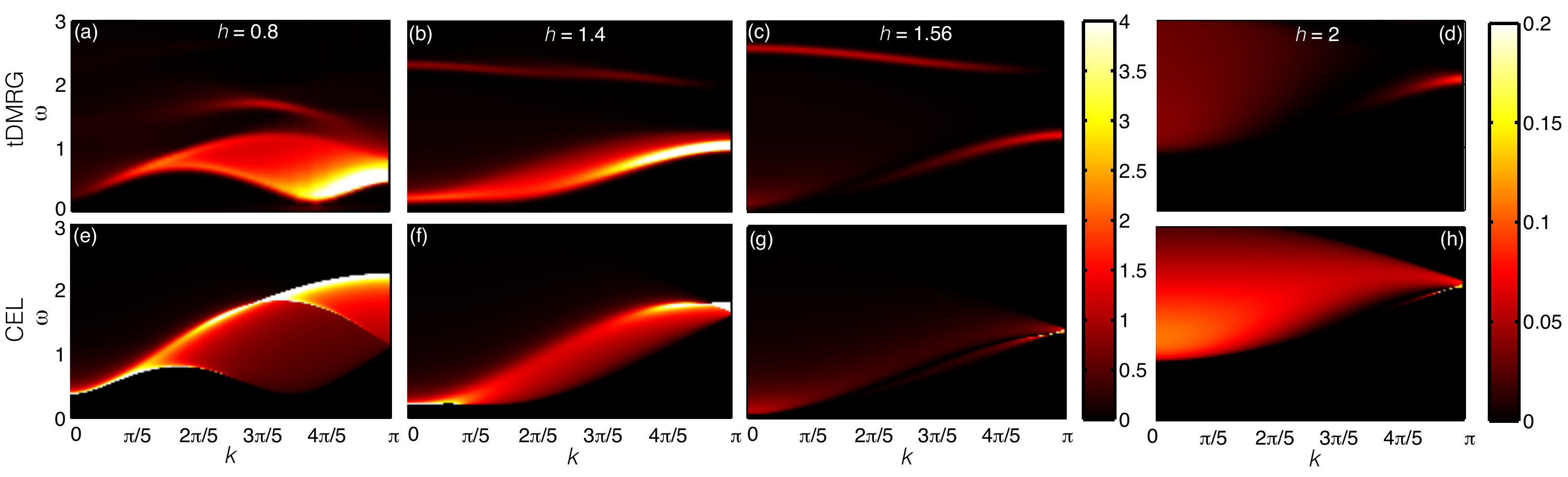}
\vspace{-15pt}
\caption{
Dynamic spin structure factor $\mathcal{S}^{xx}(k,\omega)$ longitudinal to the applied magnetic field $h$ at zero temperature. (a)-(d) show the numerical results of tDMRG [$N=100$, $t_{\rm max}=60$; see Appendix \ref{app:tDMRG_T0} for details] whereas (e)-(h) display the corresponding CEL approximation. In the spin-flop phase, (a),(b),(e),(f), the spectra show weight at low energies located at an incommensurate 
wave vector which moves towards $k=0$ at criticality, (c),(g). The weight of the spectra (d),(h) substantially decreases within the spin-polarized phase [note the different color scales]. Again, we find that the agreement between the CEL approximation and tDMRG is improving with increasing field strength.
}
\label{fig:DSSF_XX}
\end{figure*}

\subsubsection{Transverse dynamic structure factor $\mathcal{S}^{zz}(k,\omega)$}

The results for the dynamic structure factor $\mathcal{S}^{zz}(k,\omega)$ transverse to the applied magnetic field but longitudinal to the hard axis are shown in 
\Fig{fig:DSSF_ZZ}. The panels in the first row [\Figs{fig:DSSF_ZZ}(a)-(d)] illustrate our numerical tDMRG calculations, to be compared with the CEL approximation in the panels  shown in the second row [\Figs{fig:DSSF_ZZ}(e)-(h)]. The spectra in the spin-flop phase [\Figs{fig:DSSF_ZZ}(a),(b),(e),(f)] display an incoherent continuum with a gap. The majority of the spectral weight is distributed around $k=\pi$ for $h = 0.8$, but is partly shifted to  $k=0$ as the field strength is increased. At the critical point [\Figs{fig:DSSF_ZZ}(c) and (g)], the spectrum becomes gapless at the wave vector $k=\pi$ and is dominated by a single coherent mode, which remains a persistent feature also in the spin-polarized phase [\Fig{fig:DSSF_ZZ}(d),(h)] where the gap opens up again.

This coherent mode is fully captured within the CEL approximation. It possesses a dispersion of the form \cite{Caux2003}
\begin{align}
\omega(k) = \tilde J_+ \sqrt{(\cos k + \tilde h)^2 + \gamma^2 \sin^2 k} \, ,
\end{align}
where the parameters $\tilde J_+$, $\tilde h$, and $\gamma$ depend on the magnetic field $h$ and obey self-consistent mean-field equations. This dispersion is also shown as a dashed line in \Fig{fig:DSSF_ZZ}(d) with excellent agreement with the tDMRG numerics. 
At large fields, the magnetization is already close to saturation and the coherent mode essentially corresponds to a single spin-flip excitation.

As expected, the agreement between the tDMRG and the CEL approximation deteriorates with decreasing field. Interestingly, below the critical field even pronounced qualitative differences emerge. At $h=1.4$ within the spin-flop phase but close to the critical point [\Figs{fig:DSSF_ZZ}(b) and (f)], the CEL approximation still captures the low-energy branch qualitatively but it fails to describe the additional branch at higher energies, $\omega > 2$. This higher-energy branch is a distinct feature 
that is quasi coherent and possesses only a weak dispersion. It might arise from repulsively bound two-particle states that we will further discuss in Sec.~\ref{sec:conclusion}.

For even smaller fields, strong deviations between tDMRG and CEL are expected, because the latter is no longer able to describe the low-energy properties of the system, as we have already seen in the study of the order parameter in \Sec{sec:phasediag}. For a field 
$h=0.8$ [\Figs{fig:DSSF_ZZ}(a) and (e)], the higher-energy features visible around $k=0$ in the CEL spectra appear to be shifted to $k=\pi$ in the tDMRG data. At the same time, the spectral weight around $k=0$ at low energies is not captured by the CEL approximation.

\subsubsection{Longitudinal dynamic structure factor $\mathcal{S}^{xx}(k,\omega)$}

The component $\mathcal{S}^{xx}(k,\omega)$ of the dynamic spin structure factor longitudinal to the applied field is shown in \Fig{fig:DSSF_XX}. Within the CEL approximation this quantity is related to a density-density correlation function of Jordan-Wigner fermions.

Both the CEL approximation and the tDMRG calculations show that these longitudinal correlations are basically incoherent for any value of the applied magnetic field. Moreover, we find that the correlators exhibit an incommensurable low-energy feature in the spin-flop phase [\Figs{fig:DSSF_XX}(a),(b),(e),(f)], reminiscent of the incommensurability of the isotropic XY model in a longitudinal field \cite{mikeska2004one}. The incommensurable wave vector is located near $k=\pi$ (not shown) at small magnetic fields and moves towards $k=0$ at the quantum phase transition. The incommensurability becomes most apparent in \Figs{fig:DSSF_XX}(a) and (e)  for $h = 0.8$, where the wave vector corresponds to $k \approx 0.8 \pi$.
This incommensurate low-energy feature is also captured by the CEL approximation, whereas the low-energy branch at $k=\pi$ and the higher-energy excitations again substantially deviate from the tDMRG results at $h = 0.8$. For increasing field, the spectral weight decreases and becomes very small within the spin-polarized phase for all momenta as the magnetization approaches full saturation, which is illustrated by the reduced intensity of $\mathcal{S}^{xx}(k,\omega)$ in  \Figs{fig:DSSF_XX} (c),(d),(g),(h) [note that their color bars differ]. 
Similar to the transverse component in \Fig{fig:DSSF_ZZ}(b), the longitudinal component also exhibits a higher-energy branch in panel \Figs{fig:DSSF_XX}(b) and (c) that is not captured within the CEL approximation.

\begin{figure}[b]
\centering
\includegraphics[width=\linewidth]{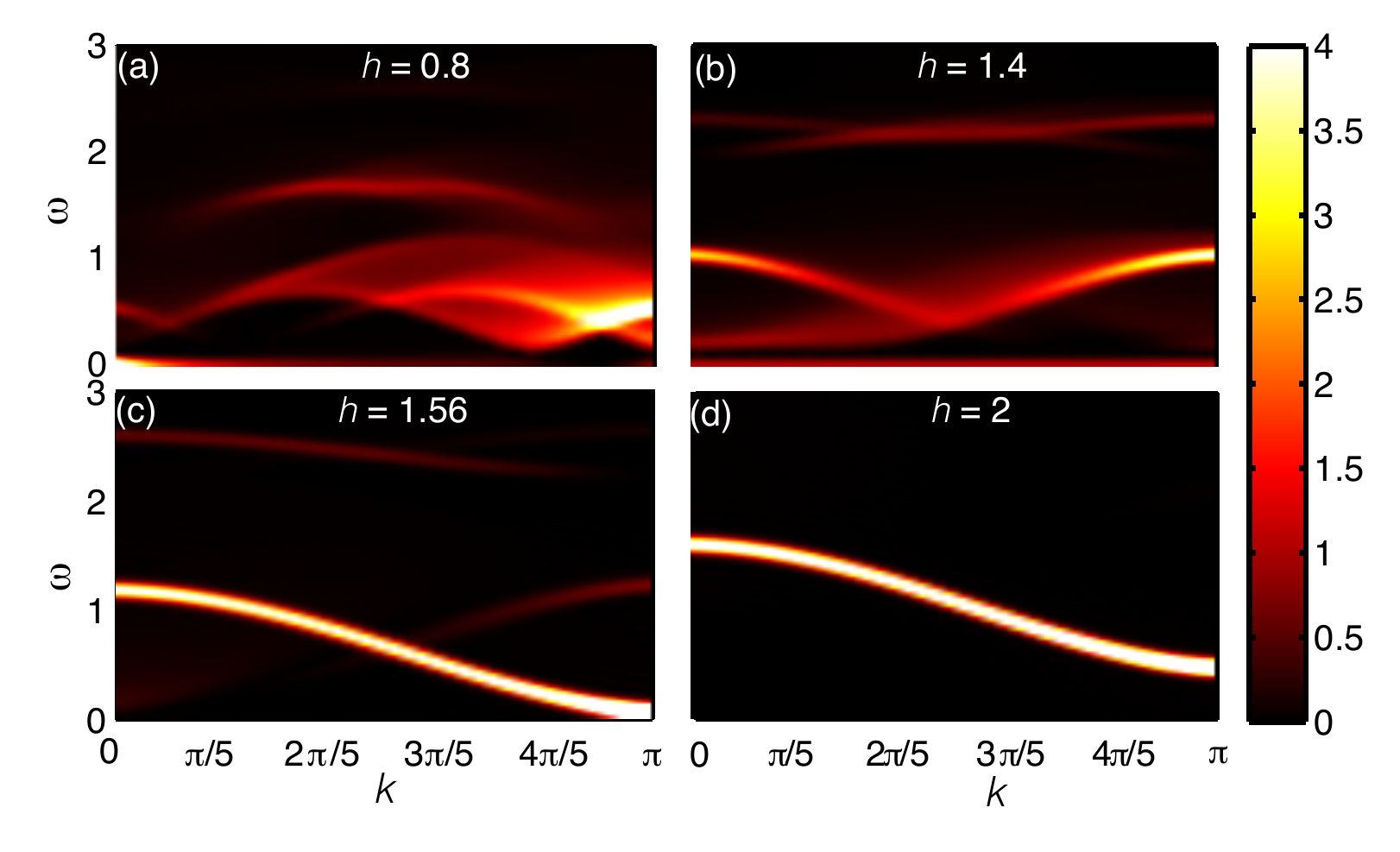}
\vspace{-25pt}
\caption{Dynamic spin structure factor $\mathcal{S}^{+-}(k,\omega)$ at $T=0$ obtained with tDMRG [$N=100$, $t_{\rm max}=60$; see Appendix \ref{app:tDMRG_T0} for details] . The spectra show the system in (a),(b) the spin-flop phase, (c) at the quantum phase transition, and (d) the spin-polarized phase.}
\label{fig:DSSF_PM}
\end{figure}

\begin{figure*}[t]
\centering
\includegraphics[width=\linewidth]{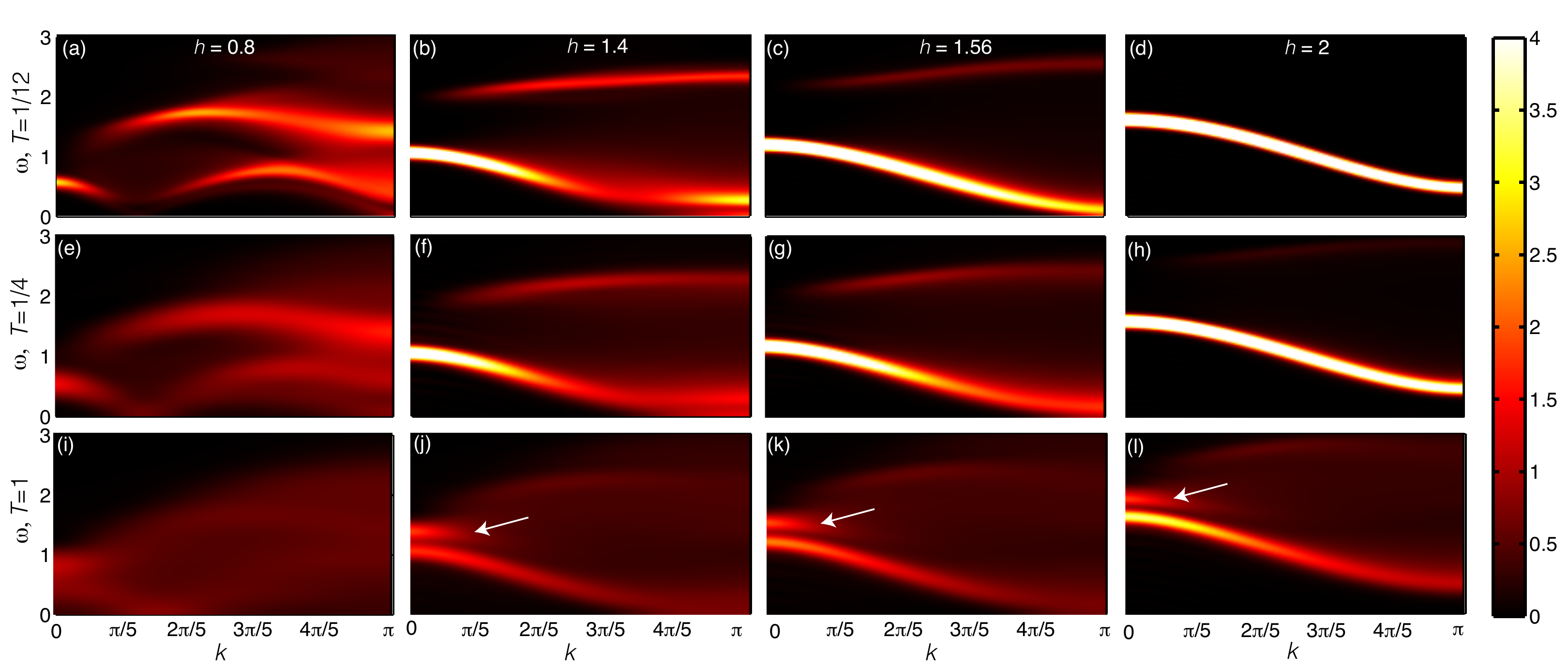}
\vspace{-15pt}
\caption{Dynamic structure factor $\mathcal{S}^{zz}(k,\omega,T)$ transverse to the applied magnetic field for three finite temperatures: (a)-(d) $T=\frac{1}{12}$, (e)-(h) $T=\frac{1}{4}$, and (i)-(l) $T=1$ obtained with tDMRG [$N=50$-$70$, $t_{\rm max}=20$-$60$; see Appendix \ref{app:tDMRG_T} for details]. Analogously to \Fig{fig:DSSF_ZZ},  the spectra show the system in the spin-flop phase (first two columns), at the quantum phase transition (third column), and in the spin-polarized phase (forth column). At finite $T$ an additional feature appears close to $k = 0$, as indicated by the white arrow.}
\label{fig:DSSF_ZZ_T}
\end{figure*}

\subsubsection{Spin-flip dynamic structure factor $\mathcal{S}^{+-}(k,\omega)$}

In \Fig{fig:DSSF_PM} we show tDMRG results for the spin-flip component $\mathcal{S}^{+-}(k,\omega)$ of spin operators within the easy plane. At high fields, the spectra are dominate by the coherent single-particle spectrum like the one of the transverse correlator in \Fig{fig:DSSF_ZZ}. In the spin-flop phase at lower fields in \Fig{fig:DSSF_PM}(a) and (b) we find that most spectral weight is distributed around $k=\pi$.

\subsection{Dynamic structure factor at finite $T$} \label{sec:DSSF_T}

We now present an analysis of the temperature dependence of the dynamical structure factor limiting ourselves, however, to a discussion of the transverse component $\mathcal{S}^{zz}(k,\omega,T)$ only. The results are obtained using real-time evolution in combination with matrix-product purification \cite{Verstraete2004,Feiguin2005}, where an auxiliary copy of the physical Hilbert space is introduced, which adopts the role of a heat bath and effectively doubles the system size. Starting from a product state consisting of maximally entangled pairs of physical and auxiliary sites, we imaginary-time evolve the system from $T=\infty$ to the desired temperature to obtain the thermal initial state $|\psi_T\rangle$ for the real-time evolution. For numerical details we refer to Appendix \ref{app:tDMRG_T}.

Considering the same field values  $h = 0.8, 1.4, 1.56, 2$ as in the previous section, we compute  $\mathcal{S}^{zz}(k,\omega,T)$ at three different temperatures $T=1,\frac{1}{4},\frac{1}{12}$, measured in units of $J$ with $k_B = 1$.
For these temperatures, the approximation of an effective spin-$\frac{1}{2}$ description for \Cs \ is still justified:
the energy gap between the doublets of the original spin-$\frac{3}{2}$ due to the single-ion anisotropy is $\Delta E \approx 4.6$ in the same units \cite{Breunig2013} so that $T \ll \Delta E$.

\Fig{fig:DSSF_ZZ_T} displays the numerical results for $\mathcal{S}^{zz}(k,\omega,T)$. First, we notice that thermal fluctuations quickly lead to a blurring of the excitation gap in the spin-flop phase [\Figs{fig:DSSF_ZZ_T} left two columns]. Already at very small temperatures  $\frac{1}{12}$ [\Figs{fig:DSSF_ZZ_T}(a) and (b)], we observe additional spectral weight being distributed around $k\approx \pi$ at $\omega = 0$. Increasing temperature further, the two spectra in the spin-flop phase show quite different behavior. Deep in the spin-flop phase for $h = 0.8$ and $T=\frac{1}{4}$ [\Fig{fig:DSSF_ZZ_T}(e)], the gap is also washed out around $k\approx 0.2 \pi$ and a lot of spectral weight is distributed towards lower energies $\omega$. At high temperatures $T=1$ [\Fig{fig:DSSF_ZZ_T}(i)], almost all spectral structures are already washed out. Closer to the phase transition at  $h = 1.4$ [\Fig{fig:DSSF_ZZ_T}(f)], the growing thermal fluctuation predominantly shift spectral weight into the region between the low and higher energy branch. For $T=1$ [\Fig{fig:DSSF_ZZ_T}(j)], the two branches have dissolved into a continuum around $k=\pi$, while the gap at $k=0$ still remains very pronounced. Interestingly, an additional spectral feature seems to arise close to $k=0$ at slightly higher energies than the low-energy branch as indicated by the arrow. We will offer an interpretation for it in the next section.

At the phase transition, thermal fluctuations cause some interesting new features in the spectrum. First of all, we note that 
the higher-energy branch becomes more  pronounced at finite $T$ while it was barely visible at $T=0$, \Fig{fig:DSSF_ZZ}(c), and not captured at all within the CEL approximation.
We also find that thermal fluctuations strongly redistribute spectral weight between the two branches for increasing temperatures. Moreover, the additional spectral feature close to $k=0$ found in \Figs{fig:DSSF_ZZ_T}(j) also appears at the quantum phase transition at $h_c \approx 1.56 $ [\Fig{fig:DSSF_ZZ_T}(k)].

A finite temperature plays only a minor role in the spin-polarized phase at $h = 2$ because the large 
excitation gap of $\Delta_e \approx 0.5$ suppresses most thermal fluctuations for $T<\Delta_e $ [\Figs{fig:DSSF_ZZ_T}(d) and (h)]. Only for temperatures above the gap, we observe thermal broadening and again the appearance of an additional excitation mode around $k=0$ [white arrow in \Fig{fig:DSSF_ZZ_T}(l)].


\section{Discussion}
\label{sec:conclusion}

In this work, we performed an extensive numerical study of the dynamic structure factor $\mathcal{S}^{\alpha \beta}(k,\omega)$ of the spin-$\frac{1}{2}$ XXZ model \eqref{eq:H12} in a transverse field for a particular value of easy-plane anisotropy $\Delta = 0.12$. Employing matrix-product-state calculations, we computed the components of the structure factor at zero and finite temperatures for various values of the transverse field with a particular focus on the Ising quantum phase transition separating a gapped spin-flop phase and a gapped spin-polarized phase at a critical dimensionless field $h_c \approx 1.56$. 

Comparing with previous approximate analytical calculations of Caux, Essler, and L\"ow (CEL) \cite{Caux2003} for $T=0$, we confirmed that at large fields, $h \gtrsim h_c$, the correlator $\mathcal{S}^{zz}$ transverse to the applied field is governed by a coherent single-particle mode, which in the large-field limit basically corresponds to a single spin-flip excitation of the almost polarized chain. At smaller fields, $\mathcal{S}^{zz}$ looses coherence as it becomes dominated by the two-particle continuum. The correlator $\mathcal{S}^{xx}$ longitudinal to the field, on the other hand, is mostly incoherent. 

Our numerical study has revealed two distinct features in the dynamic structure factor that deserve special attention: $(i)$ an additional  relatively sharp mode located at higher energies $\omega > 2 J$ clearly visible in $\mathcal{S}^{zz}$, see \Fig{fig:DSSF_ZZ}(b), as well as in $\mathcal{S}^{xx}$, see \Fig{fig:DSSF_XX}(b) and (c); and $(ii)$ additional weight emerging at finite temperature just above the low-energy branch close to zero wave vector, see white arrow in \Fig{fig:DSSF_ZZ_T}. In particular, a mode at higher energies $\omega > 2 J$ is not anticipated within the mean-field approximation of CEL. 

Let us speculate about the origin of these additional features. A possible candidate for $(i)$ the higher-energy mode is a repulsively bound two-particle state, which goes beyond the mean-field approximation. The presence of such a bound state is at least supported from an analysis in the large field limit. In this limit, the ground state is completely polarized and a particle excitation just corresponds to a single spin flip. While a spin flip loses Zeeman energy, $J h$, it gains twice the bond energy $J/2$ due to the antiferromagnetic alignment with its neighboring spins. This also applies for each spin flip of the two-particle excitation provided that they are separated by at least two sites. If spin flips occupy adjacent sites, however, they gain only half of the bond energy giving rise to an effective repulsive interaction $J$. In lowest order in $1/h$, this repulsion gives rise to a bound state above the two-particle continuum similar to the doublon in the Hubbard model \cite{Winkler2006}. At high fields, its weight is probably too small to be observable in the dynamic structure factor, but it might survive at smaller fields, giving rise to the signatures observed in our spectra. The lifetime of this repulsively bound state could be large at small temperatures, as its decay requires the interaction with additional particles in order to release its energy \cite{Strohmaier2010,Sensarma2010,Chudnovskiy2012}.
At finite temperatures, the  thermal occupation of particles will facilitate the decay, which might explain the fading of the higher-energy mode in the spectra in \Fig{fig:DSSF_ZZ_T} with increasing $T$. It is striking that the signature $(ii)$ close to zero wave vector in the transverse dynamic structure factor $\mathcal{S}^{zz}$  identified by the white arrow in \Fig{fig:DSSF_ZZ_T}, which is reminiscent of a Villain mode \cite{Villain1975}, gains weight with the simultaneous vanishing of the higher-energy mode. It is therefore tempting to speculate that this feature $(ii)$ is associated with the decay of the repulsively bound pair. 

It might be worth exploring these spectral features further in future theoretical 
work. The physics of repulsively bound particle pairs should be particularly transparent in the Ising limit of the XXZ spin-$\frac{1}{2}$ chain for a longitudinal field close to its triple point \cite{Trippe2010}. On the experimental side, the dynamic structure factor  considered in this work might be observable with the help of inelastic neutron scattering experiments on the compound \Cs. This material thus offers the opportunity to study the rich structure of the dynamic spin-spin correlations of the XXZ spin-$\frac{1}{2}$ chain in a regime where it is not integrable with interesting effects emerging already on the two-particle level. We hope that our study motivates such experiments in the near future.


\acknowledgments
M.G. acknowledges helpful discussions and an earlier collaboration with O. Breunig and T. Lorenz that motivated this work. B.B. thanks the Tensor-Network group at UCI for their hospitality and in particular S. R.~White for pointing out the concept of smooth boundary conditions. This research was supported by the Deutsche Forschungsgemeinschaft through the Excellence Cluster ``Nanosystems Initiative Munich'', SFB/TR~12, SFB~631 and  WE4819/2-1 (A.W.). 
\\


\appendix

\section*{Appendix: Numerical details} \label{app:numerics}
This appendix discusses the numerical methods used to obtain the results presented in the main part of the paper.  \Sec{app:tDMRG} deals with  tDMRG, \Sec{app:CheMPS} with CheMPS, and \Sec{app:CheMPSvstDMRG} offers a detailed comparison of their efficiency within the context of the present spin-$\frac{1}{2}$ XXZ model.

\subsection{tDMRG}  \label{app:tDMRG}
This section elaborates on the details of our tDMRG implementation employed to generate the results for the dynamic structure factor at zero and finite temperatures in the main part of this work. 

\subsubsection{Zero temperature}  \label{app:tDMRG_T0}
To evaluate the zero-temperature structure factor by means of real-time evolution, we first have to determine the time-dependent ground-state correlators
\begin{equation} \label{tdmrg:dssf}
\mathcal{S}^{\alpha \beta}(j,t) = e^{iE_0t} \langle \psi_0 | \hat{S}^{\alpha}_j e^{-i\hat{H}t} \hat{S}^{\beta}_0 | \psi_0 \rangle
\end{equation}
 for various times $t$ and distances $j$. To this end, we initialize the ground state $|\psi_0 \rangle$ in terms of an MPS employing DMRG 
 \cite{White1992,schollwock_2005,schollwock_2011} before applying the local perturbation $\hat{S}^{\beta}_0$ in the middle of the chain (labeled with $j_M=0$) to generate  $| \phi \rangle =  \hat{S}^{\beta}_0 | \psi_0 \rangle$. $| \phi \rangle $  is the initial state for the real-time evolution, $| \phi (t) \rangle = e^{-i \hat{H}t} |\phi\rangle$, which is carried out using standard tDMRG techniques \cite{Feiguin_PRL_2004,Vidal_PRL_2004,Daley_2004,schollwock_2011}. This amounts to splitting the time-evolution operator $e^{-i \hat{H}t}$ into a product of $M$ small time steps $\tau = t/M$. For systems with short-ranged interactions, each term $e^{-i \hat{H}\tau}$ is decomposed into a product of local operator via a Suzuki-Trotter decomposition. For Hamiltonians with nearest-neighbor interactions only,  such as (\ref{eq:H12}), this results in combining all interaction terms corresponding to even and odd numbered bonds respectively, i.e, $\hat{H} = \hat{H}_e + \hat{H}_o$. Note that all terms in one group commute with each other but the terms in $\hat H_e$ generally do not commute with the ones in $\hat H_o$. The second-order Suzuki-Trotter decomposition for the time-evolution operator then reads 
 \begin{equation} 
 e^{-i \hat{H}\tau} = e^{-i \hat{H}_e\tau/2} e^{-i \hat{H}_o\tau} e^{-i \hat{H}_e\tau/2} + \mathcal{O}(\tau^3) \, .
 \end{equation} 
 The time evolution is carried out by repeatedly applying the Trotter-decomposed evolution operator to the initial state $|\phi\rangle$. For every (or a subset of) time step(s) we evaluate the two-point correlators $\mathcal{S}^{\alpha \beta}(j,t)$ for all possible values of $j$ on the finite chain. In the end, we compute the Fourier transform in time $t$ and real space $j$ to obtain the dynamic structure factor of Eq.~\eqref{eq:DSSF_def}.
 
 Such calculations are typically affected by two major error sources: 
 \begin{enumerate}[(1)]
 \item The Trotter decomposition introduces an error of the order $ \mathcal{O}(\tau^3)$ because it ignores the noncommutativity of odd and even terms of the Hamiltonian. This so-called Trotter error can be dealt with by using a higher-order decomposition \cite{Trotter} or a smaller time step. 
 \item The spreading of the excitation over time causes a growth of entanglement in the state during the time evolution, which typically requires the bond dimension of the MPS to increase exponentially towards longer time scales. This effectively restricts the accessible time scale to some maximum time $t_{\rm max}$, the value of which strongly depends on the specific model and parameter regime. 
 \end{enumerate}
 
The finite-time limit also puts a constraint on the resolution of the spectral functions in frequency space. In order to remove artificial finite-time oscillations in the spectra, one needs to include some type of broadening when performing the Fourier transform to  frequency space. Here we choose to include a Gaussian filter $\exp[-\eta^2 t^2]$ in the time integral in \Eq{eq:DSSF_def} and choose $\eta$ dependent on $t_{\rm max}$. Hence, the resulting spectral functions contain the exact spectral features convolved with a Gaussian $\exp[ -\omega^2/(2 W^2)]$, with a frequency resolution $W=\sqrt{2} \eta$. In some cases, linear prediction can be used to avoid the artificial broadening and extract more spectral information from the time series \cite{White_PRB_2008_LinPre,Barthel_PRB_2009}. We refrain from employing linear prediction in this work, because we found its results were very sensitive to changes of the regularization parameter and the statistical window on the given time scale for the present model.

The zero-temperature tDMRG calculations in \Sec{sec:DSSF} were performed on a chain with open boundary conditions and $N=100$ spins, which is large enough to prevent any finite-size reflections for the considered time scales. We worked with a second-order Suzuki-Trotter decomposition and used a time step $\tau=0.05$, which is small enough in the context of the present model that the Trotter error becomes negligible. Moreover, the bond dimension $D$ of the time-evolved MPS $|\phi(t)\rangle$ was chosen adaptively by keeping all singular values larger than $\epsilon_{\rm SVD} =10^{-4}$ during the application of the Trotter gates. We stopped the time evolution at $t_{\rm max} = 60$ and worked with a broadening parameter $\eta=0.033$, which corresponds to an energy resolution of $W = \sqrt{2}\eta = 0.047 J$ using the Gaussian filter of \Ref{Feiguin_PRL_2004} in the reconstruction of the dynamic structure factor. In practice, this lead to a maximum bond dimension of $D<1400$ during the last time step. Furthermore, we note that a setup with smooth boundary conditions, employed in \Sec{sec:phasediag} to minimize finite-size effects in static quantities, is not particularly well suited for dynamic calculations. Its decreasing energy scales at the chain's ends introduce a set of low-energy states, which significantly alter the entanglement growth during time evolution.

\subsubsection{Finite temperatures} \label{app:tDMRG_T}

The above approach can be generalized with minor modifications to calculate finite-temperature correlators
\begin{equation}
\mathcal{S}^{\alpha \beta}(j,t,T) = \langle \psi_T | e^{i\hat{H}t} \hat{S}^{\alpha}_j e^{-i\hat{H}t} \hat{S}^{\beta}_0 | \psi_T \rangle \, .
\end{equation}
In this case, the local perturbation $\hat{S}^{\beta}_0$ is no longer applied to the ground state $|\psi_0\rangle$ but rather to a thermal state $| \psi_T \rangle$, which either represents the purified density matrix \cite{Verstraete2004} or one state of an ensemble of minimally entangled typical thermal states (METTS) \cite{White09,Stoudenmire10}, depending on the chosen finite-temperature algorithm. Since the evolution operator acting on the bra cannot be factored out as a phase factor anymore, one has to carry out two independent real-time evolutions, $|\phi(t) \rangle =  e^{-i\hat{H}t} \hat{S}^{\beta}_0 | \psi_T \rangle$ and $ |\Phi (t)\rangle=e^{-i\hat{H}t} | \psi_T \rangle$ and evaluate $\mathcal{S}^{\alpha \beta}(j,t,T) = \langle \Phi (t) |\hat{S}^{\alpha}_j |\phi(t) \rangle$ accordingly.

The finite-temperature tDMRG calculations in \Sec{sec:DSSF_T} were performed in the purification setup on an open chain of $N=$ 50-70 physical spins (corresponding to a total number of $N_{\rm tot } =$ 100-140 sites in the purified scheme), where the time scales were again chosen such that no finite-size reflections occurred. We set  $\epsilon_{\rm svd} = 10^{-4}, \, 10^{-5}$ during the real- and imaginary-time evolution, respectively, and chose a Trotter step of $\tau=0.05$ in both cases. Since the entanglement of the MPS during time evolution grows much more rapidly the higher the temperature, the accessible time scale varied between $t_{\rm max} = 60$ for $T=1/12$ and  $t_{\rm max} = $ 20-$40$ for $T=1$.\footnote{In order to reach these time scales, we applied a backward time evolution on the auxiliary states for $T=1,\frac{1}{4}$, which significantly reduced the growth of entanglement \cite{Karrasch12}. Note that we refrained from exploiting time-translation invariance to reach even larger times \cite{Barthel13}, since it would have required to carry out tDMRG runs individually for each distance $j$.} Although thermal broadening dominates at high temperatures on the considered time scale, we nevertheless included a broadening parameter $\eta=0.05$ in the Fourier transform for consistency.

\subsection{CheMPS} 
\label{app:CheMPS}
In this section we discuss the basics of CheMPS, which are relevant for the detailed comparison to tDMRG in Appendix \ref{app:CheMPSvstDMRG}.

With CheMPS we are able to work directly  in frequency space and compute dynamic correlators of the type
 \begin{equation} \label{eq:GF}
\mathcal{S}^{\alpha\beta} (j,\omega) = \langle \psi_0 | \hat{S}_j^{\alpha} \delta(\omega-\hat{H}+E_0) \hat{S}_0^{\beta} | \psi_0 \rangle \, .
\end{equation} 
 The CheMPS approach expands the $\delta$-function in \Eq{eq:GF} in terms of Chebyhsev polynomials of the first kind, $T_n$. To ensure the convergence of the Chebyshev expansion, the Hamiltonian has to be rescaled such that its support is fully contained in the interval $[-1,1]$. One way to achieve this is to use a linear mapping $\hat{H'} = (\hat{H}-E_0)/a - b$, $\omega' = \omega/a - b$ with the two rescaling factors $a, b$ chosen properly.
 
  \Ref{Wolf_PRB_2015} showed that the details of the rescaling procedure clearly affect the efficiency of the calculation.  It is usually most efficient to map the support of the spectral function close to the lower boundary of the interval $[-1,1]$, where the zeros of the individual Chebyshev polynomials are densely distributed. This can be achieved by using a ``$b=1$'' setup, which is in the following distinguished from the ``$b=0$'' setup, where the support of the spectral function lies at the center of  $[-1,1]$.

After proper rescaling, the correlator in \Eq{eq:GF} can be represented with Chebyhsev coefficients 
\begin{equation}
\mu_n(j) = \langle \psi_0 | \hat{S}_j^{\alpha} T_n(H') \hat{S}_0^{\beta} | \psi_0 \rangle, 
\end{equation}
leading to
\begin{equation} \label{eq:A_CheMPS}
\mathcal{S}^{\alpha\beta} (j,\omega)  = \frac{1}{a} \sum_{n=0}^{N_{\rm Che}} w_n(\omega') \mu_n(j) T_n (\omega'),
\end{equation}
with  $w_n(\omega) = (2-\delta_{n0})/(\pi\sqrt{1-\omega^2})$. The numerical demanding part is to determine the Chebyshev coefficients $\mu_n(j)$. To this end, one employs standard MPS techniques and exploits the recursion relations of the Chebyshev polynomials to iteratively generate the Chebyshev vectors
\begin{eqnarray}
|t_n\rangle &=& 2\hat{H}' |t_{n-1}\rangle - |t_{n-2}\rangle, \label{eq:CheRec}\\
|t_0\rangle &=& \hat{S}^{\beta}_0 | \psi_0 \rangle, \quad |t_1\rangle = \hat{H}' |t_0\rangle.
\end{eqnarray}
Thus by storing only three MPS per expansion step, we can iteratively evaluate the Chebyshev coefficients $\mu_n(j)$ by computing overlaps of the type $\mu_{n}(j) = \langle \psi_0 | \hat{S}^{\alpha}_j |  t_n \rangle$ for all values of $j$ on the finite chain. Analogous to real-time evolution, it is typically more convenient to carry out the Fourier transform from real- to momentum-space after completing the expansion, instead of applying momentum-space operator $\hat{S}^{\beta}_k$ to the starting state. In this way, only a single calculation is required to obtain the spectrum at various momenta. Moreover, a local perturbation  $\hat{S}^{\beta}_0$ leads to a significantly reduced entanglement growth during the expansion. 

The increase of entanglement stored in $|t_n\rangle$ at higher expansion orders is caused by the repeated application of the Hamiltonian $\hat{H}$ to the MPS and is necessary from a physical point of view to represent the spreading of the local excitation in real space over time. This results in a roughly exponentially growing demand on the numerical resources in order to store and manipulate Chebyshev vectors. Therefore, the expansion is limited to some finite order $N_{\rm Che}$, at which the computational costs ``hit the exponential wall''. The finite-order cut off introduces numerical artifacts in the dynamic correlators, which can be removed by including coefficients $g_n$ of a broadening kernel in \Eq{eq:A_CheMPS}, which smears out the higher order terms and generate a smooth spectrum. Alternatively, it is also possible to determine the full resolvent function in \Eq{eq:GF} for a nonzero value of $\eta$ \cite{braun} or, in some cases, to avoid broadening at all by means of linear prediction \cite{Ganahl_PRB_2014}.

Recently, \Ref{Tiegel_PRB_2014} expanded CheMPS to  determine spectra also at finite temperatures. To this end, they formulated the Chebyshev expansion in terms of a Liouvillian and  matrix-product purification. It is also possible to combine CheMPS with METTS, but for technical reasons this turned out to be very inefficient \cite{SYMETTS}.

\subsection{tDMRG vs. CheMPS} \label{app:CheMPSvstDMRG}

In the following, we compare the numerical efficiency of the two methods, tDMRG and CheMPS. CheMPS has been frequently applied in practice \cite{Thomale_PRB_2013,Ganahl_PRB_2014,Tiegel_PRB_2014,Wolf_PRB_2014,braun,Halimeh15,Rausch15,Tiegel15}, but no conclusive answer has yet been presented to the question whether it provides a computationally more efficient framework over real-time evolution to simulate spectral functions. Whereas tackling this question in full generality would go beyond the scope of this work, we present below a brief analysis of the efficiency of CheMPS in the present context.

Our main conclusion is that CheMPS produces zero-temperature results of similar quality as tDMRG at comparable computational costs. Accordingly, the CheMPS setup, too, needs to appropriately deal with a growing amount of entanglement in the MPS to produce reliable results.

In order to compare real-time evolution and CheMPS, we have studied the  spin-$\frac{1}{2}$ XXZ chain Hamiltonian \eqref{eq:H12}
with $N=100$ spins directly at quantum criticality $h = 1.56$ and $T=0$. Starting by placing an excitation in the middle of the chain, we take $\hat{S}^{\beta}_0 | \psi_0 \rangle$ as the initial state for both the real-time evolution and the Chebyshev expansion. The CheMPS simulation is carried out in two setups: one with $b=0$ in the linear mapping, see Appendix~\ref{app:CheMPS}, and $N_{\rm Che}= 4800$ iterations, 
 another  with  $b=0.995$ and $N_{\rm Che}= 2100$ iterations. The reference tDMRG calculation uses the data from \Fig{fig:DSSF_ZZ}(c). As previously, we adapt the bond dimension of the MPS by truncating according to $\epsilon_{\rm SVD} = 10^{-4}$ in every Trotter step as well as any Chebyshev iteration \Eq{eq:CheRec} during the entire calculation.

\begin{figure}[b]
\centering
\includegraphics[width=\linewidth]{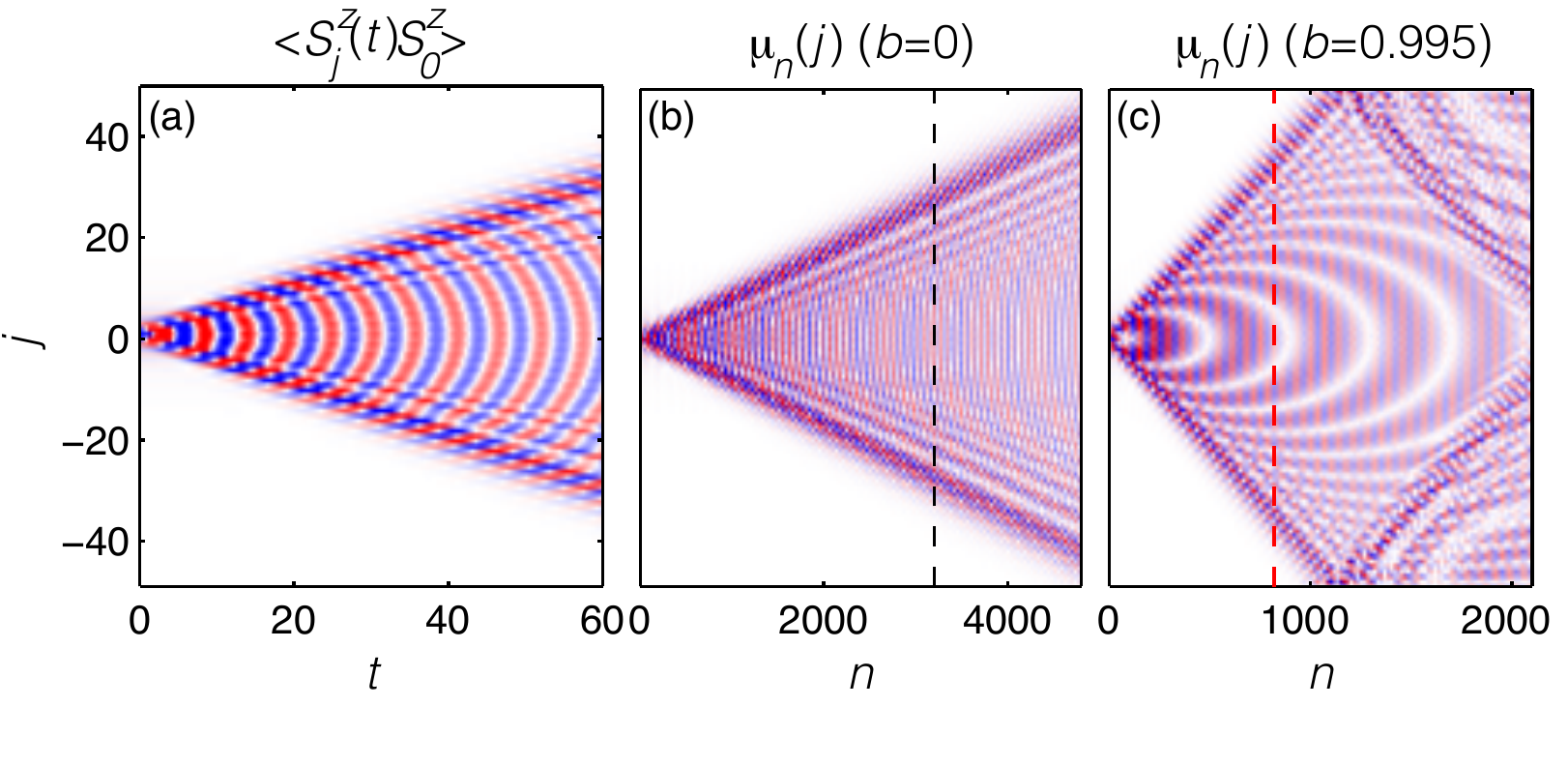}
\vspace{-10pt}
\caption{Evolution of the excitation over (a) time $\langle \hat{S}^z_j \hat{S}^z_0(t) \rangle$ and (b),(c) iteration order $\mu_n(j) = \langle \psi_0 | \hat{S}^{\alpha}_j |  t_n \rangle$. }
\label{fig:SSF_CheMPSvstDMRG}
\end{figure} 

\Fig{fig:SSF_CheMPSvstDMRG} displays the corresponding evolution of the excitation with time, $\langle \hat{S}^z_j (t) \hat{S}^z_0 \rangle$, and iteration order, $\mu_n(j) = \langle \psi_0 | \hat{S}^{\alpha}_j |  t_n \rangle$, respectively. In all cases, the initially localized excitation spreads out in real-space showing the typical light-cone structure. We clearly observe that finite-size reflections are not present up to the maximum time $t_{\rm max}=60$ in the tDMRG simulation [\Fig{fig:SSF_CheMPSvstDMRG}(a)]. The same applies to the CheMPS results of the $b=0$ setup in \Fig{fig:SSF_CheMPSvstDMRG}(b). Following the literature, the final iteration corresponds to an effective time scale $t\sim N_{\rm Che}/a \approx 60$, which is equivalent to the maximum time of the tDMRG reference calculation. However, the excitation in (b) is already spread out significantly further in the system than at the end of the tDMRG calculation. This deviation becomes even more apparent studying the $b=0.995$ setup in (c), which in principle should evolve according to the same effective time scale as the rescaling factor $a$ is unchanged. In reality, the excitation has already reached the boundary of the system after $n\approx 1100$ iterations. Reflections at both boundaries become strongly visible for higher iterations. This suggests that the effective time scale of $t^* = 60$ is already reached significantly sooner in the $b=0.995$ setup, which is in agreement with the findings of \Ref{Wolf_PRB_2015}. 

\begin{figure}[t]
\centering
\includegraphics[width=\linewidth]{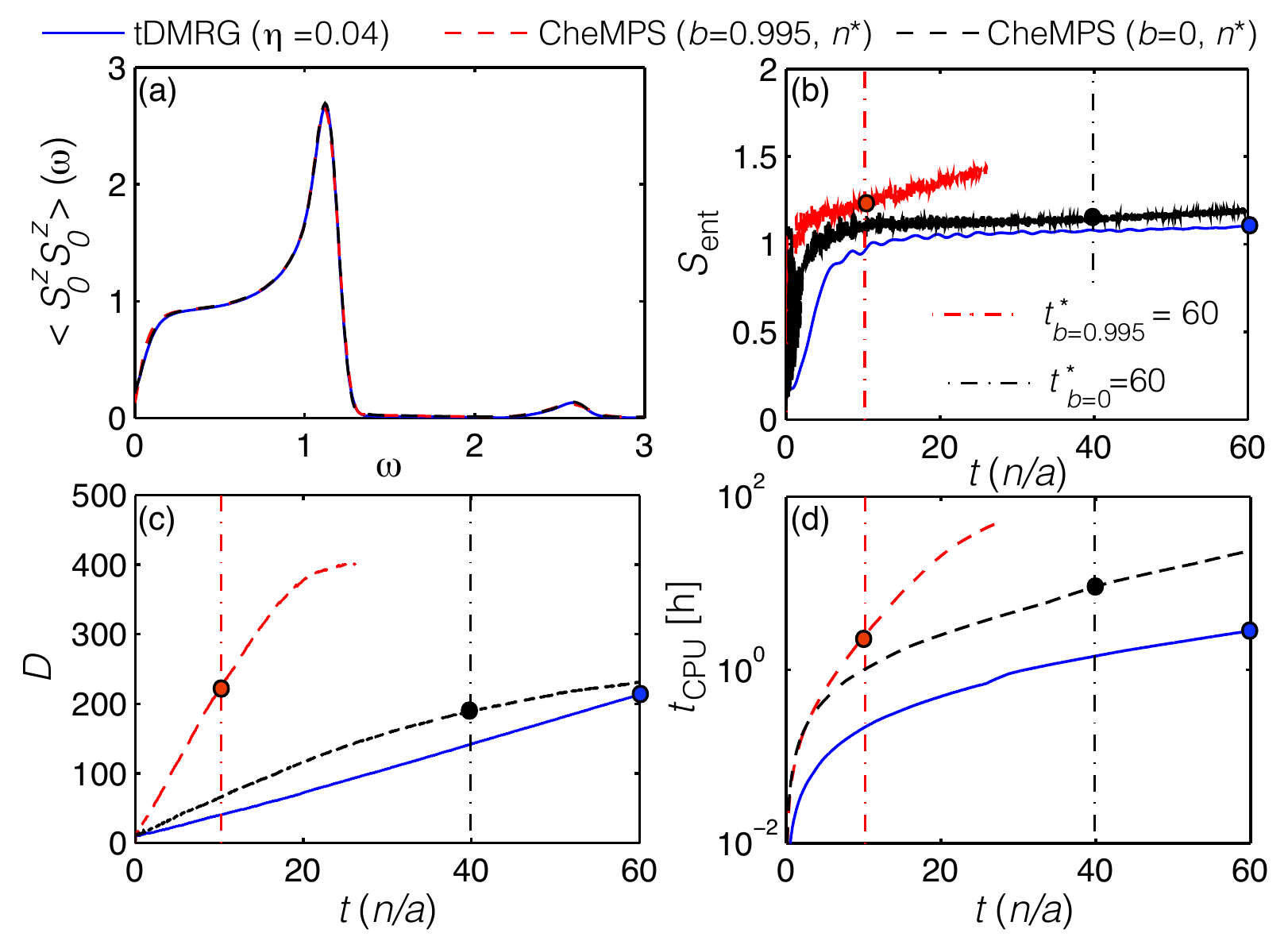}
\vspace{-10pt}
\caption{(a) Local spectral function $\langle \hat{S}^z_0 \hat{S}^z_0 \rangle (\omega)$ obtained from tDMRG and CheMPS for the spin-$\frac{1}{2}$ model  for \Cs \,  with $N=100$ spins directly at the phase boundary for $h = 1.56$ and $T=0$. (b)-(d) Comparison of entanglement entropy $S_{\rm ent}$, bond dimension $D$, and cumulative CPU time $t_{\rm CPU}$. }
\label{fig:CheMPSvstDMRG}
\end{figure}

Hence, we conclude that only $n^* < at_{\rm max}$ CheMPS iterations have to be carried out in order to obtain spectral data with comparable accuracy as in the reference tDMRG simulation. This is illustrated in \Fig{fig:CheMPSvstDMRG}(a), where the local spectral function $\langle \hat{S}^z_0 \hat{S}^z_0 \rangle (\omega)$ obtained from tDMRG and CheMPS data is displayed. We use only the first $n^*$ moments of the respective CheMPS calculation and a Jackson kernel in the Chebyshev reconstruction to mimic both the maximum time cut off and the Gaussian broadening in the Fourier transform of the real-time data, choosing $n^*$ such that the agreement with the reference data is best. These iterations $n^*$ are indicated by the dashed vertical lines in \Figs{fig:SSF_CheMPSvstDMRG}(b) and (c). As one would intuitively expect, the excitation is spread over  approximately the same distance after these $n^*$ iterations as in the tDMRG calculation at $t_{\rm max}$.

Thus we can restrict our efficiency analysis to the first $n \leq n^*$ iterations in order to conduct a reasonable comparison to tDMRG. \Figs{fig:CheMPSvstDMRG}(b)-(d) show the entanglement entropy, bond dimension and accumulated CPU time, respectively. The tDMRG data is plotted in real-time units $t$, whereas the CheMPS results are displayed with a rescaled iteration number $n/a$ for better comparability. Again, the dashed vertical lines indicate the iteration  $n^*/a$  of interest. First of all, we note that the Chebyshev vectors at $n^*$ in both setups are slightly more entangled than the time-evolved MPS [\Fig{fig:CheMPSvstDMRG}(b)], although this is not reflected in the respective bond dimensions at $n^*$ or $t_{\rm max}$, respectively: The final time-evolved MPS has a bond dimension $D=213$, the corresponding Chebyshev vectors in the $b=0$ and $b=0.995$ setup carry a somewhat comparable number of many-body states ($D=188$ and $D=218$, respectively). This indicates that both methods require very similar amounts of numerical resources in order to reproduce the same spectral information. A comparison of CPU times further confirms this, as tDMRG and  $b=0.995$ CheMPS require almost the identical amount of total CPU-time, namely $t_{\rm CPU} = 2.8$ hours on a 8-core machine. The CheMPS calculation in the $b=0$ setup takes approximately three times longer due to the larger number of iterations necessary to reach the same time scale. 

We have conducted this study only for a single model and set of parameters, thus we cannot provide an unambiguous answer to whether a spectral function is best represented in terms of Fourier modes or Chebyshev functions. However, we learned here that both methods are affected by the dynamical entanglement growth in a very similar matter. Therefore, it seems rather unlikely that one method can significantly outperform the other. For this reason, we have only applied one approach, namely tDMRG, to generate the results presented in Secs.~\ref{sec:DSSF} and \ref{sec:DSSF_T}. Our analysis would have to be extended to other parameters and systems in order to give a fully conclusive answer. For instance, we expect that tDMRG outperforms CheMPS at finite $T$, since (i) the Liouvillian formulation of CheMPS requires a factor $a$ twice as large as in the $T=0$ setup; (ii) the more efficient $b=1$ setup, which aims to shift the support of the spectral function close to the lower boundary of the rescaled interval $[-1,1]$, might not be appropriate if finite temperatures shift the support to higher energies; (iii) there exists no counterpart to time-translation invariance, which  allows us to effectively double the maximum time scale in the tDMRG setup \cite{Barthel13}. On the other hand, CheMPS might be the preferred choice for zero-temperature calculations in models with long-ranged interactions, where a Trotter-based time evolution is no longer feasible.


\FloatBarrier


\begin{thebibliography}{99}

\bibitem{Algra1976}
H. Algra, L. de Jongh, H. Bl\"ote, W. Huiskamp, and R. Carlin, Physica (Amsterdam) B+C {\bf 82}, 239 (1976).

\bibitem{McElearney1977}
J. N. McElearney, S. Merchant, G. E. Shankle, and R. L. Carlin, J. Chem. Phys. {\bf 66}, 450 (1977).

\bibitem{Smit1979}
J. Smit and L. De Jongh, Physica (Amsterdam) B+C {\bf 97}, 224 (1979).

\bibitem{Duxbury1981}
P. M. Duxbury, J. Oitmaa, M. N. Barber, A. van der Bilt,
K. O. Joung, and R. L. Carlin, Phys. Rev. B {\bf 24}, 5149 (1981).

\bibitem{Kenzelmann2002}
M. Kenzelmann, R. Coldea, D. A. Tennant, D. Visser, M. Hofmann, P. Smeibidl, and Z. Tylczynski, Phys. Rev. B
{\bf 65}, 144432 (2002).


\bibitem{Chatterjee2003}
I. Chatterjee, J. Magn. Magn. Mater. {\bf 265}, 363 (2003).

\bibitem{Siahatgar2008}
M. Siahatgar and A. Langari, 
Phys. Rev. B {\bf 77}, 054435 (2008).

\bibitem{Breunig2013}
O. Breunig, M. Garst, E. Sela, B. Buldmann, P. Becker, L. Bohaty, R. M\"uller, and T. Lorenz,
Phys. Rev. Lett. {\bf 111}, 187202 (2013).

\bibitem{Breunig2015}
O. Breunig, M. Garst, A. Rosch, E. Sela, B. Buldmann, P. Becker, L. Bohaty, R. M\"uller, and T. Lorenz,
Phys. Rev. B {\bf 91}, 024423 (2015).

\bibitem{Toskovic2016}
R. Toskovic, R. van den Berg, A. Spinelli, I. S. Eliens, B. van den Toorn, B. Bryant, J.-S. Caux, and A. F. Otte, Nat. Phys. {\bf 12}, 656--660 (2016).



 

\bibitem{Caux2003}
J.-S. Caux, F. H. L. Essler, and U. L\"ow,
Phys. Rev. B {\bf 68}, 134431 (2003).



\bibitem{Feiguin_PRL_2004}
S. R. White and A. E. Feiguin,
Phys. Rev. Lett. {\bf 93}, 076401 (2004).


\bibitem{Vidal_PRL_2004}
G. Vidal,
Phys. Rev. Lett. {\bf 93}, 040502 (2004).

\bibitem{Daley_2004}
A. J. Daley, C. Kollath, U. Schollw\"ock, and G. Vidal,
J. Stat. Mech.: Theor. Exp. (2004) P04005.



\bibitem{Verstraete2004}
F. Verstraete, J. J. Garcia-Ripoll, and J. I. Cirac,
Phys. Rev. Lett. {\bf 93}, 207204 (2004).

\bibitem{Feiguin2005}
A. E. Feiguin and S. R. White,
Phys. Rev. B {\bf 72}, 220401(R) (2005).

\bibitem{Zhe2015}
Z. Wang, J. Wu, S. Xu, W. Yang, C. Wu, A. K. Bera, A. T. M. Nazmul Islam, B. Lake, D. Kamenskyi, P. Gogoi, H. Engelkamp, A. Loidl, J. Deisenhofer,
arXiv:1512.01753 [cond-mat.str-el].

\bibitem{holzner}
A. Holzner, A. Weichselbaum, I. P. McCulloch, U. Schollw\"ock, and J. von Delft, 
Phys. Rev. B {\bf 83}, 195115 (2011).

%


\bibitem{White1992}
S. R. White,
Phys. Rev. Lett. {\bf 69} (19), 2863 (1992).

\bibitem{schollwock_2005}
U.~Schollw\"ock,
\RMP {\bf 77}, 259 (2005).

\bibitem{schollwock_2011}
U.~Schollw\"ock,
Ann. Phys. {\bf 326}, 96 (2011).



\bibitem{braun}
A. Braun and P. Schmitteckert,
Phys. Rev. B {\bf 90}, 165112 (2014).


\bibitem{Ganahl_PRB_2014}
M. Ganahl, P. Thunstr\"om, F. Verstraete, K. Held, and H. G. Evertz,
Phys. Rev. B {\bf 90}, 045144 (2014).

\bibitem{Wolf_PRB_2014}
F. A. Wolf, I. P. McCulloch, O. Parcollet, and U. Schollw\"ock, 
Phys. Rev. B {\bf 90}, 115124 (2014).

\bibitem{Tiegel_PRB_2014}
A. C. Tiegel, S. R. Manmana, T. Pruschke, and A. Honecker,
Phys. Rev. B {\bf 90}, 060406 (2014).


\bibitem{Ramasesha_1997}
S. Ramasesha, S. K. Pati, H. Krishnamurthy, Z. Shuai, and J. Br\`edas,
Synth. Met. {\bf 85}, 1019 (1997).

\bibitem{Kuehner_PRB_1999}
T. D. K\"uhner and S. R. White, 
Phys. Rev. B {\bf 60}, 335 (1999).

\bibitem{Jeckelmann_PRB_2002}
E. Jeckelmann, 
Phys. Rev. B {\bf 66}, 045114 (2002).

\bibitem{Jeckelmann_2008}
E. Jeckelmann, 
Prog. Theor. Phys. Suppl. {\bf 176}, 143 (2008).

\bibitem{Zhu_2013}
Z. Zhu, D. A. Huse, and S. R. White, 
Phys. Rev. Lett. {\bf 110}, 127205 (2013).

\bibitem{Zhu_2013b}
Z. Zhu, D. A. Huse, and S. R. White, 
Phys. Rev. Lett. {\bf 111}, 257201 (2013).


\bibitem{Vekic1993}
M. Veki\'{c} and S. R. White, 
Phys. Rev. Lett. {\bf 71}, 4283 (1993).

\bibitem{Vekic1996}
M. Veki\'{c} and S. R. White,
Phys. Rev. B {\bf 53}, 14552 (1996).

\bibitem{Caux2005}
J.-S. Caux and J. M. Maillet, Phys. Rev. Lett. {\bf 95}, 077201  (2005).

\bibitem{Pereira2006}
R. G. Pereira, J. Sirker, J.-S. Caux, R. Hagemans, J. M. Maillet, S. R. White, and I. Affleck,
Phys. Rev. Lett. {\bf 96}, 257202 (2006).

\bibitem{mikeska2004one}
H.-J. Mikeska and A.~K. Kolezhuk,
{\it One-dimensional magnetism}, Lect. Notes Phys. {\bf 645}, 1--83, (Springer Berlin Heidelberg, 2004).


\bibitem{Winkler2006}
K. Winkler, G. Thalhammer, F. Lang, R. Grimm, J. Hecker Denschlag, A. J. Daley, A. Kantian, H. P. B\"uchler, and P. Zoller,
Nature (London) {\bf 441}, 853-856 (2006).

\bibitem{Strohmaier2010}
N. Strohmaier, D. Greif, R. J\"ordens, L. Tarruell, H. Moritz, T. Esslinger, R. Sensarma, D. Pekker, E. Altman, and E. Demler, 
Phys. Rev. Lett. {\bf 104}, 080401 (2010).


\bibitem{Sensarma2010}
R. Sensarma, D. Pekker, E. Altman, E. Demler, N. Strohmaier, D. Greif, R. J\"ordens, L. Tarruell, H. Moritz, and T. Esslinger, 
Phys. Rev. B {\bf 82}, 224302 (2010).

\bibitem{Chudnovskiy2012}
A. L. Chudnovskiy, D. M. Gangardt, and A. Kamenev,
Phys. Rev. Lett. {\bf 108}, 085302 (2012).


\bibitem{Villain1975}
J. Villain,
Physica B {\bf 79}, 1 (1975). 

\bibitem{Trippe2010}
C. Trippe, F. G\"ohmann, and A. Kl\"umper, 
J. Stat. Mech. {\bf 2010}, P01021.


\bibitem{Trotter}
N. Hatano and M. Suzuki,
\emph{Quantum Annealing and Other
Optimization Methods}, edited by A. Das and B.K. Chakrabarti
(Springer, Berlin, 2005), pp. 37--68.

\bibitem{White_PRB_2008_LinPre}
S. R. White and I. Affleck,
Phys. Rev. B {\bf 77}, 134437 (2008).

\bibitem{Barthel_PRB_2009}
T. Barthel, U. Schollw\"ock, and S. R. White, 
Phys. Rev. B {\bf 79}, 245101 (2009).

\bibitem{White09}
S. R. White,
Phys. Rev. Lett. {\bf 102}, 190601 (2009).

\bibitem{Stoudenmire10}
M. E. Stoudenmire and S. R. White,
New J. Phys. {\bf 12}, 055026 (2010).

\bibitem{Wolf_PRB_2015}
F. A. Wolf, J. A. Justiniano, I. P. McCulloch, and U. Schollw\"ock, 
Phys. Rev. B {\bf 91}, 115144 (2015).

\bibitem{SYMETTS}
B. Bruognolo, J. von Delft, and A. Weichselbaum,
Phys. Rev. B {\bf 92}, 115105 (2015).

\bibitem{Karrasch12}
C. Karrasch, J. H. Bardarson, and J. E. Moore,
Phys. Rev. Lett. {\bf 108}, 227206 (2012).

\bibitem{Barthel13}
T. Barthel,
New J. Phys. {\bf 15}, 073010 (2013).


\bibitem{Thomale_PRB_2013}
R. Thomale, S. Rachel, and P. Schmitteckert,
Phys. Rev. B {\bf 88}, 161103 (2013).

\bibitem{Halimeh15}
J. C. Halimeh, F. Kolley, I. P. McCulloch,
Phys. Rev. B {\bf 92}, 115130 (2015).

\bibitem{Tiegel15}
A. C. Tiegel, A. Honecker, T. Pruschke, A. Ponomaryov, S. A. Zvyagin, R. Feyerherm, and S. R. Manmana,
Phys. Rev. B {\bf 93}, 104411 (2016).

\bibitem{Rausch15}
R. Rausch and M. Potthoff,
New J. Phys. {\bf 18}, 023033 (2016).



\end{thebibliography}
\end{document}